\documentclass[prl,twocolumn,amsmath,amssymb,english,superscriptaddress]{revtex4-1}

\usepackage[dvipdfm]{graphicx}
\usepackage{natbib}
\usepackage{subfigure}
\usepackage{tabularx}
\usepackage{epsfig}
\usepackage{longtable}
\usepackage{amsfonts}
\usepackage{rotating}
\usepackage{subfigure}

\usepackage{hyperref}
\usepackage{babel}

\usepackage{currfile}

\usepackage{color}

\newcommand{\bea}{\begin{eqnarray}}
\newcommand{\eea}{\end{eqnarray}}
\newcommand{\la}{\label}
\newcommand{\be}{\begin{equation}}
\newcommand{\ee}{\end{equation}}



\begin{document}

\title{Topological Zero-Energy Modes in Gapless Commensurate Aubry-Andr\'e-Harper Models.}
 
\author{Sriram Ganeshan}
\affiliation{Condensed Matter Theory Center and Joint Quantum Institute, Department of Physics, University of Maryland, College Park, MD 20742, USA}

\author{Kai Sun}
\affiliation{Condensed Matter Theory Center and Joint Quantum Institute, Department of Physics, University of Maryland, College Park, MD 20742, USA}
\affiliation{Department of Physics, University of Michigan, Ann Arbor, MI 48109, USA}

\author{S. Das Sarma}
\affiliation{Condensed Matter Theory Center and Joint Quantum Institute, Department of Physics, University of Maryland, College Park, MD 20742, USA}

\date{\today}

%%%%%%%%%%%%%
\begin{abstract}

%Recent optical experiments have unravelled a novel topological classification of quasicrystals. Theoretically, this topological feature appears as a function of previously known but often discarded phase term in the Aubry-Andre (AAH) model also known as Harper model.  In this work we identify a different class of topological edge modes in the half flux state of the off-diagonal Harper model. This novel topology is shown to be connected to the Majorana physics. We will demonstrate simple experimental settings that can realize this interesting feature in a bosonic framework.
 The Aubry-Andr\'e or Harper (AAH) model has been the subject of extensive theoretical research in the context of quantum localization. Recently, it was shown that one-dimensional quasicrystals described by the {\it incommensurate} AAH model has non-trivial topology. In this paper,  we show that the {\it commensurate off-diagonal} AAH model is topologically nontrivial in the gapless regime and supports zero-energy edge modes. Unlike the incommensurate case, the nontrivial topology in the off-diagonal AAH model is attributed to the topological properties of the one-dimensional Majorana chain.  We discuss the feasibility of experimental observability of our predicted topological phase in the {\it commensurate} AAH model.
%\setcounter{tocdepth}{1}
%\tableofcontents

\end{abstract}
%%%%%%%%%%%%%

\maketitle

%\tableofcontents

\textit{Introduction.}
Anderson localization is a quantum-interference-induced disorder-tuned  quantum phase transition on a tight binding lattice where the system wavefunction changes from being extended (``metal") to exponentially localized (``insulator") at a critical value of the disorder strength \cite{Anderson}.  In one-dimensional (1D) systems, Anderson localization is trivial since the critical disorder is zero, and all states for any finite disorder are localized.  The absence of a true quantum phase transition makes 1D Anderson localization rather uninteresting from the perspective of the physics of disorder-tuned metal-insulator transition.
%One of the most important concepts that emerged from the studies of disorder effects is the idea of Anderson localization\cite{Anderson}. Anderson localization is an effect where quantum interference induces a transition by turning the extended Bloch wavefunctions into localized states in the presence of random disorders. Although this mechanism has been proposed more than half a century ago, the direct observation of this phenomenon remains a challenge in solid state systems, mainly due to the lack of direct control over impurities.  
%
 However, Aubry and Andr\'e predicted the existence of a localization transition for certain 1D quasiperiodic systems akin to Harper model \cite{AA,harper55}, where the transition arises from the existence of an incommensurate potential of finite strength mimicking disorder in a 1D tight binding model. The form of this quasiperiodic potential is usually chosen to be a cosine function  incommensurate with the underlying periodic tight-binding 1D lattice.   This result has led to extensive theoretical studies of disorder effects in the AAH model during the last few decades~\cite{sankarprl88, thoulessprl88, sankarprb90, biddlepra09, biddleprl10}. Recent experiments~\cite{lahini2009, kraus1} have realized the quasiperiodic  AAH model in optical lattices and observed the signature of a localization transition~\cite{lahini2009} in agreement with theory~\cite{AA}.

 One very interesting aspect of the 1D AAH model is that it can be exactly mapped to the 2D Hofstadter model~ \cite{harper55, hofstadter}. The Hosftadter model describes the topologically-nontrivial 2D quantum Hall (QH) system on a lattice~\cite{laughlin,tknn,Avron,hatsugai}.  This mapping implies that the 1D AAH model must have topologically-protected edge states similar to the gapless edge states of the QH effect. Recently, these edge states have been observed experimentally \cite{kraus1} and this mapping has been used to topologically classify 1D quasicrystals described by the {\it incommensurate} AAH model~\cite{kraus1, kraus2,langprl12, lang12}. 
 %Topological classification of quasi-crystals (QC)\cite{kraus1,kraus2} and their realization in optical experiments has brought together three seemingly different areas of research in condensed matter systems. In Ref.~\onlinecite{kraus1}, a specific 1D QC was considered which is described by Aubry-Andre (AAH) model (Harper equation). AAH model was  experimentally realized via quasiperiodic lattice of coupled single-mode waveguides. %An underlying topological classification was unearthed by viewing energy eigenstates as a function of the phase parameter that corresponds to higher dimensional degree of freedom. This new perspective has provided a topological framework to the AAH class of 1D Hamiltonians.   Theoretically, a complete topological classification of 1D quasicrystal of AAH type and Fibonacci QC was achieved by mapping the generalized Harper equation to parent 2D tight binding lattice in presence of a magnetic flux. 
%
In this letter, we study the {\it commensurate off-diagonal} AAH model. We note in the passing that although we use the AAH nomenclature to discuss our model (mainly to establish connection with existing work in the literature), our proposed commensurate off-diagonal system is simply a `bichromatic' 1D system with two competing underlying 1D commensurate periodic potentials with arbitrary phases. 
The hopping amplitude of the off-diagonal AAH model has a cosine modulation in the real space commensurate with the lattice. This is in contrast to the diagonal AAH model which has a cosine modulation in the potential energy term. Both these versions can be unified within a generalized AAH model (also known as generalized Harper model~\cite{han94, kraus2} -~- we use the AAH terminology throughout this paper to emphasize that Aubry-Andr\'e and Harper are equivalent models for our purpose). 
 In particular, we focus on the parameter range where the AAH model is gapless and thus cannot be mapped onto a QH system, a situation which has so far been thought to be trivial and not considered at all in the vast literature on the AAH model.  Surprisingly we find that edge states exist in this seemingly topologically trivial model.  These edge states are topologically protected, but they belong to a different topological class than the QH edge. 
 The topological origin of these edge states is analogous to that of zero-energy edge states along the zigzag edge of graphene~\cite{ryu2002,halperin09,montambaux10} and is directly connected to the $Z_2$ topological index of the Kitaev model~\cite{kitaev01} and a particular case can also be mapped to the Su-Schreiffer-Heeger (SSH) model ~\cite{ssh79}. Thus our work shows a  hitherto undiscovered deep connection between the AAH model,  graphene,  the Kitaev model and the SSH model. To make the nontrivial topology transparent, we rewrite our model in the Majorana basis. It must be emphasized that the {\it commensurate off-diagonal} AAH model studied in this work has extended Bloch 1D band bulk states for all parameter values (and no localization transition at all), and our establishing its topological edge behavior differs qualitatively from all earlier recent work on topological properties of the {\it incommensurate} AAH model~\cite{kraus1,kraus2,langprl12,lang12,diptiman12,indu12}.
 %%%%%%%%%%%%%%%%%%%%%%%%%%%%%%%%%%%%%%%%%%%%%%%%%%%%%%%%%%%%%%%%%%%%%%%%%%%%%%%%%%%%%%%%
\begin{figure*}[htb!]
  \centering
  \subfigure[Topologically-nontrivial off-diagonal AAH model for $b=\frac{1}{2}$ ($\pi$-flux)]{\includegraphics[scale=0.37]{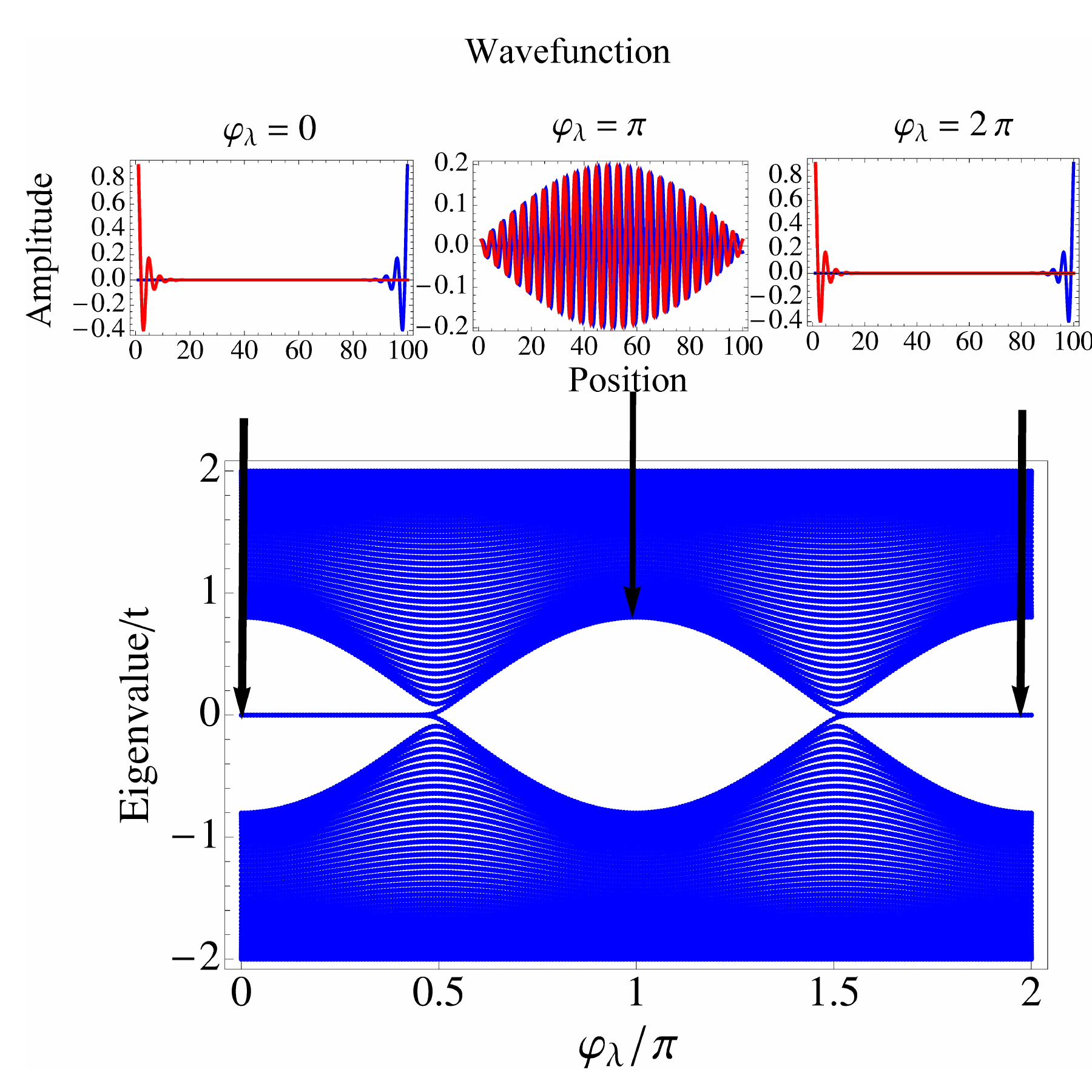}}\quad\quad
  \subfigure[Topologically trivial diagonal AAH model for $b=\frac{1}{2}$ ($\pi$-flux)]{\includegraphics[scale=0.37]{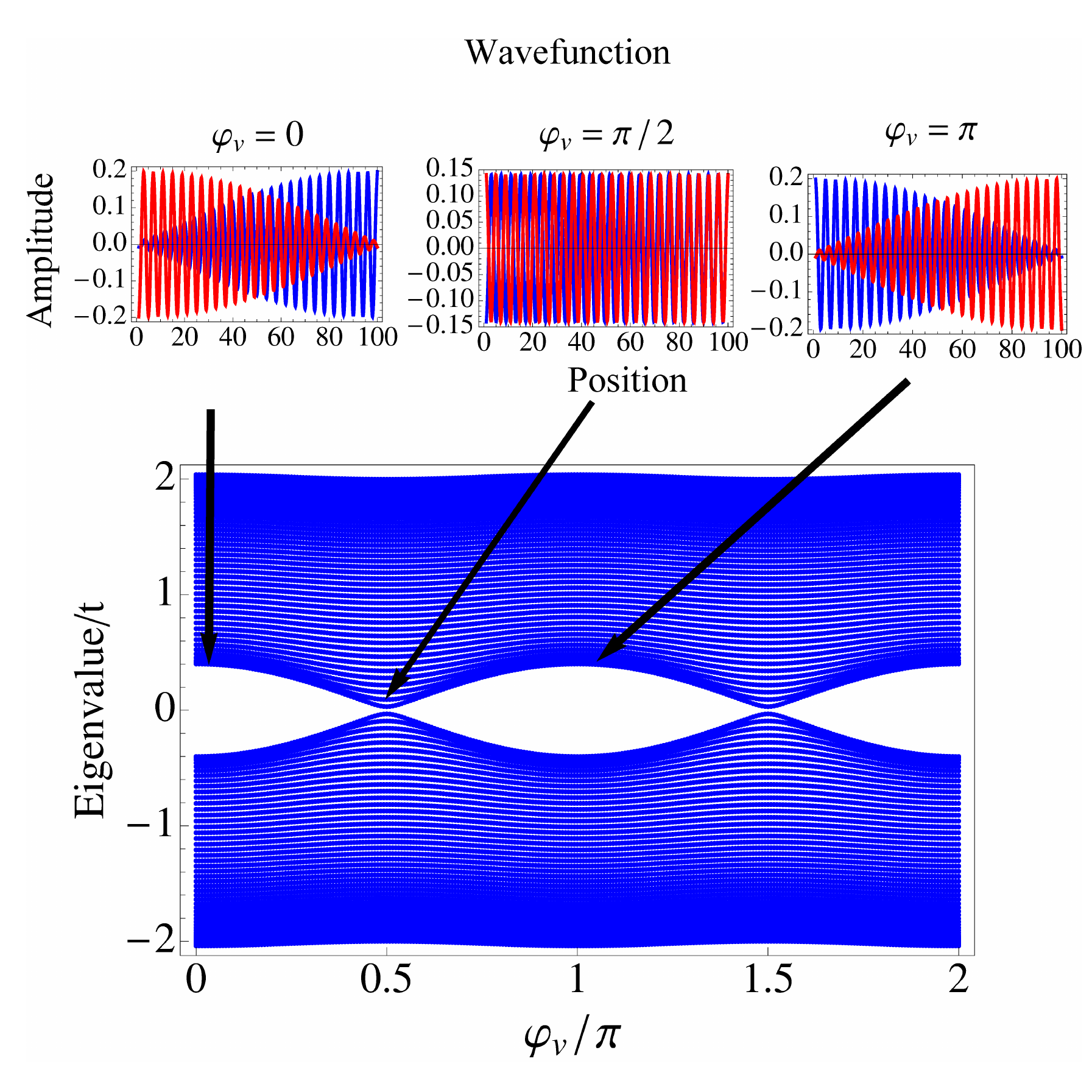}}
  \caption{Upper panel shows the two zero-energy eigenstates (red and blue lines) plotted as a function of position for three different values of $\varphi$. The lower panel is the energy spectrum plotted as a function of $\varphi$ for 100 sites. (a) Spectrum and eigenstates with parameters $t=1$, $\lambda=0.4$, $v=0$ and $b=1/2$. (b) Spectrum and eigenstates with parameters $t=1$, $v=0.4$, $\lambda=0$ and $b=1/2$.  }
  \la{energystates}
\end{figure*}
%\begin{center}
%\begin{figure*}[!]
%\includegraphics[width=6.0cm,height=7cm]{fig1a.pdf}\hspace{0.1cm}\includegraphics[width=6.0cm,height=7cm]{fig1b.pdf}
%\caption{Upper panel are the two zero mode eigenstates (red and blue lines) plotted as a function of position for three different values of $\varphi$. The lower panel is the energy spectrum plotted as a function of $\varphi$ for 100 sites and with parameters $t=1$ $\lambda^{od}=0.4$, $\lambda^{d}=t'=0$ . }
%\label{energystates}
%\end{figure*}
%\end{center}
%%%%%%%%%%%%%%%%%%%%%%%%%%%%%%%%%%%%%%%%%%%%%%%%%%%%%%%%%%%%%%%%%%%%%%%%%%%%%%%%%%%%%%%%
\vspace{-0.015cm}
In addition, we provide specific experimental setups for the realization of the off-diagonal AAH model using a 1D lattice composed of coupled single mode waveguides with varying lattice spacings or a double-well optical lattice. To understand whether the topological edge states can be observed in a real experimental setup, we examine the robustness of these edge states against next order hopping and fluctuations in lattice potentials, both of which arise in a real experimental system. Although the topological index cannot be clearly defined in the presence of these realistic effects, explicit numerical calculations indicate that these edge modes are robust and observable for a wide range of parameter space. This robustness, arising from the topological nature of the zero modes, is of crucial importance for the experimental verification of our prediction.

\textit{Model.}
We consider the generalized 1D AAH model, which is described by the following
Hamiltonian
\bea
H&=&\sum_{n=1}^{N-1} t[1+\lambda \cos(2\pi b n+\varphi_{\lambda})] c^{\dagger}_{n+1}c_n+h.c.\nonumber\\
&&+\sum_{n=1}^{N} v \cos(2\pi b n+\varphi_{v}) c^{\dagger}_{n}c_n.
\la{eq:Hamiltonian}
\eea
This 1D chain has $N$ sites ($n=1$, $2$, $\ldots$, $N$).  We adopt open boundary conditions with $n=1$ and $n=N$ being the two edge sites. To be consistent with previous literature on topological edge modes~\cite{kitaev01}, we consider {\it fermionic} particles which are created and annihilated by fermionic operators $c^\dagger_n$ and $c_n$. We emphasize, however, that our work and all conclusions are equally valid for the corresponding bosonic case since we are considering a noninteracting 1D quantum system. 
The first term in the Hamiltonian is the kinetic energy from the  nearest-neighbor hopping, and the last term describes the on-site potential energy. The inhomogeneity in hopping strength and potential energy terms are described via cosine modulations of the strength $\lambda$ and $v$ respectively. 
The cosine modulations have periodicity $1/b$  and phase factor $\varphi_{\lambda}$ and $\varphi_{v}$. The special case with $\lambda=0$ ($v=0$)  corresponds to the diagonal (off-diagonal) AAH model. The generalized AAH model can be derived starting from an ancestor 2D Hofstadter model with next-nearest-neighbor (diagonal) hopping terms~\cite{hiramoto89,han94,kraus2}.  
Starting from a 2D ancestor, the phase terms $\varphi_{\lambda}$ and $\varphi_v$ are related by $\varphi_{\lambda}=\varphi_v+\pi b$. Experimentally, one can design setups where both $\varphi_{\lambda}$  and  $\varphi_v$ can be tuned independently, so we keep our notations general with  $\varphi_{\lambda}$  and  $\varphi_v$  as independent variables. 

For irrational $b$, the diagonal AAH model  ($\lambda=0$, $v \ne 0$)  shows a localization transition as $v$ is increased beyond the critical value ($v=2t$) with all states being extended(localized) for $v < 2t~(v>2t)$~\cite{AA, kohmoto83,hiramoto89}. For rational $b$, it is known that by treating the phase  $\varphi_{v}$ as the momentum of another spatial dimension, the diagonal AAH model can be mapped onto a 2D Hofstadter lattice with $2\pi b$ magnetic flux  per plaquette~\cite{harper55,hofstadter}. For $b\ne 1/2$, the Hofstadter lattice has gapped energy bands with nontrivial topology, described by nonzero Chern numbers. Therefore, localized edge modes are expected for a finite-sized system with boundary. It is worthwhile to mention here that although the mapping to Hofstadter lattice is well-defined for rational values of $b$, the topologically-protected edge states remain stable even if $b$ takes irrational values.
Here, we start from the special case $b=1/2$ and in later part, we will show that all the conclusions can be generalized as long as $1/b$ is an even integer. For $b=1/2$, the off-diagonal AAH model can be mapped onto a 2D Hofstadter model with $\pi$ flux per plaquette. Under time-reversal transformation, a $\pi$ flux simply turns into a $-\pi$ flux. Since the magnetic flux terms are only well defined modulo $2\pi$ for a lattice, the system is then invariant under the time-reversal transformation. Therefore, the system shows no QH effect and thus has no QH edge modes. In fact, this Hofstadter lattice model has no band gap but contains two Dirac points with linear dispersion in analogy to graphene. This gapless $\pi$-flux state has been extensively studied in the context of algebraic spin liquids (see for example Refs.~~\cite{wenbook, fradkinbook} and references therein). 
We have calculated the energy spectrum for diagonal and off-diagonal AAH models at different values of $\varphi_{v}$ and $\varphi_{\lambda}$ (Fig.~\ref{energystates}). For periodic boundary conditions (not shown here), both models show two energy bands with two Dirac points. If we perform the same calculations on a 1D lattice with open edges, the energy spectrum (as a function of $\varphi_{\lambda,v}$) for the diagonal AAH model remains the same as in the case of periodic boundary conditions. But for the off-diagonal AAH model, zero-energy states are observed for $-\pi/2<\varphi_{\lambda}<\pi/2$ as shown in Fig.~\ref{energystates}. By examining the wavefunctions associated with these zero-energy states, we find that they are actually boundary states localized around the two edges of the system, as shown in the inset of Fig.~\ref{energystates}. These are the topological zero-energy (edge) modes alluded to in the title of our paper. 

\textit{Degenerate Majorana modes}. The edge states we find here have a topological origin, which can be understood analytically by rewriting the off-diagonal AAH model (for $b=1/2$) in the Majorana basis. We define $c_{2n}=\gamma_{2n}+i\tau_{2n},\  c_{2n+1}=\tau_{2n+1}+i\gamma_{2n+1}$, where $\gamma$ and $\tau$ are  two species of Majorana fermions. In this new basis, the off-diagonal AAH model becomes,
\bea
H&=&\sum_{n}[\Delta_{-}(\gamma_{2n}\gamma_{2n-1})+\Delta_{+}(\gamma_{2n}\gamma_{2n+1})]\nonumber\\
&-&\sum_{n}[\Delta_{-}(\tau_{2n}\tau_{2n-1})+\Delta_{+}(\tau_{2n}\tau_{2n+1})],
               \la{offdmajorana}
               \eea
%%%%%%%%%%%%%%%%%%%%%%%%%%%%%%%%%%%%%%%%%%%%%%%%%%%%%%%%%%%%%%%%%%%%%%%%%%%%%%%%%%%%%%%%%?
\begin{center}
\begin{figure}[htb!]
\includegraphics[width=7.5cm,height=6cm]{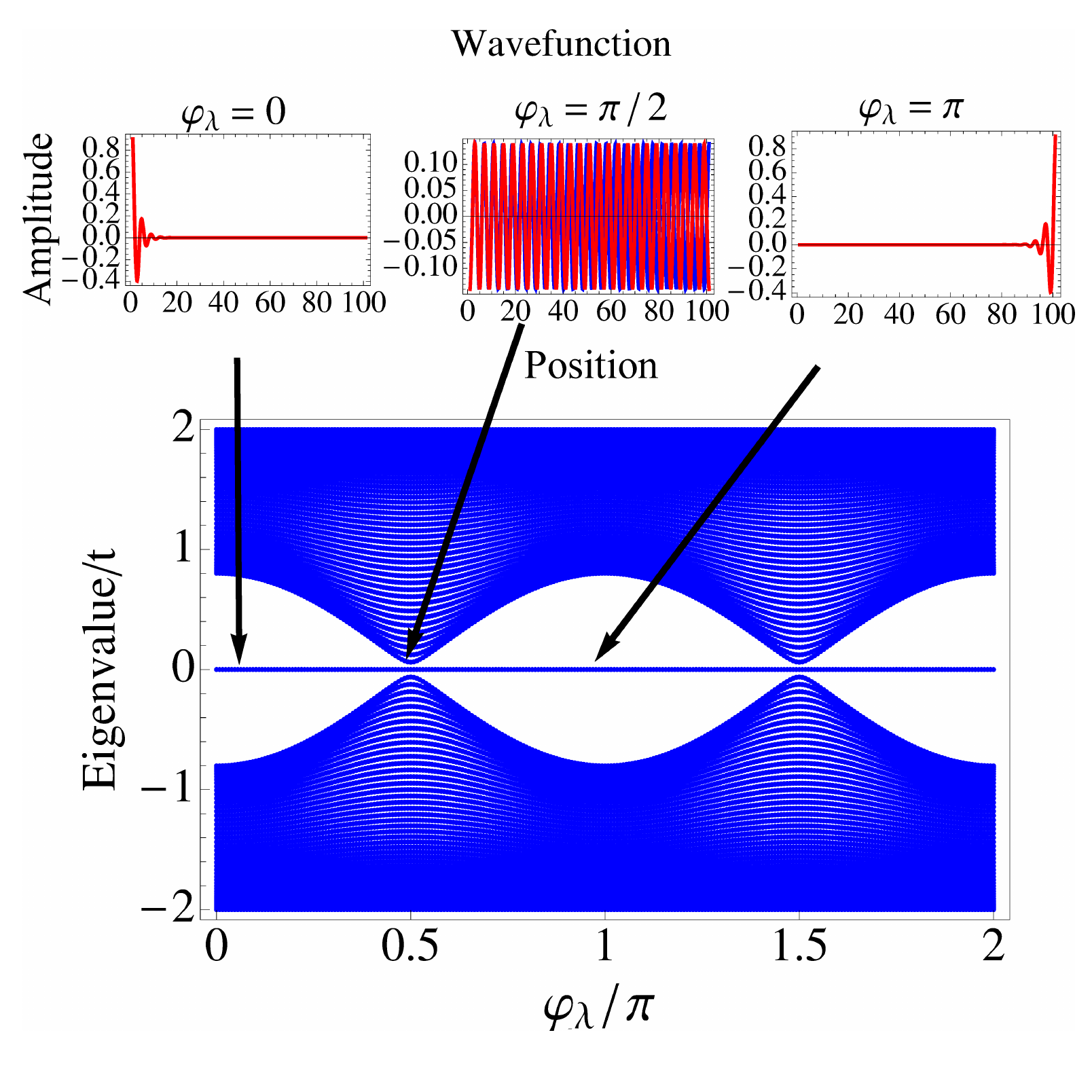}
\vspace{-0.5cm}
\caption{Energy spectrum with odd number of sites ($N=101$) with parameters $t=1$, $b=1/2$, $v=0$ and $\lambda=0.4$.  A single zero-energy mode is always localized on either one of the edges except for the Dirac points as shown by the wavefunction plots.} 
\label{odd}
\end{figure}
\end{center}
%%%%%%%%%%%%%%%%%%%%%%%%%%%%%%%%%%%%%%%%%%%%%%%%%%%%%%%%%%%%%%%%%%%%%%%%%%%%%%%%%%%%%%%%%
where $\Delta_{\pm}=2it(1\pm\lambda\cos\varphi_{\lambda})$. Here, the system contains two identical 1D Majorana chains which are decoupled from each other. 
In the study of 1D topological superconductors, it is known that a Majorana fermion chain supports a $Z_2$ topological index~\cite{kitaev01}. For $|\Delta_+|>|\Delta_-|$, the Majorana chain is topologically nontrivial and has one zero-energy Majorana mode localized at each edge. For the opposite regime $|\Delta_+|<|\Delta_-|$, the system is topologically trivial with no edge modes. For our model, this implies that two Majorana modes, which are equivalent to a Dirac edge mode, are expected at each of the two edges for $\cos\varphi_{\lambda}>0$, which agrees perfectly with the numerical results shown in Fig.~\ref{energystates}. This $Z_2$ index is in fact the parity (even or odd) of the integer topological index (Z) for a 1D system with chiral symmetry~\cite{kitaev09,ryu09}. Since we only allow for short-range (nearest-neighbor) hopping terms in our model, the integer $Z$ index can only take values $0$ or $1$ (i.e. $Z_2$). If the kinetic energy is dominated by the longer-range hopping terms, higher topological index can be achieved, which is beyond the scope of this Letter.
%%%%%%%%%%%%%%%%%%%%%%%%%%%%%%%%%%%%%%%%%%%%%%%%%%%%%%%%%%%%%%%%%%%%%%%%%%%%%%%%%%%%%%%%%?
\begin{center}
\begin{figure}[htb!]
\includegraphics[width=7.5cm,height=6.5cm]{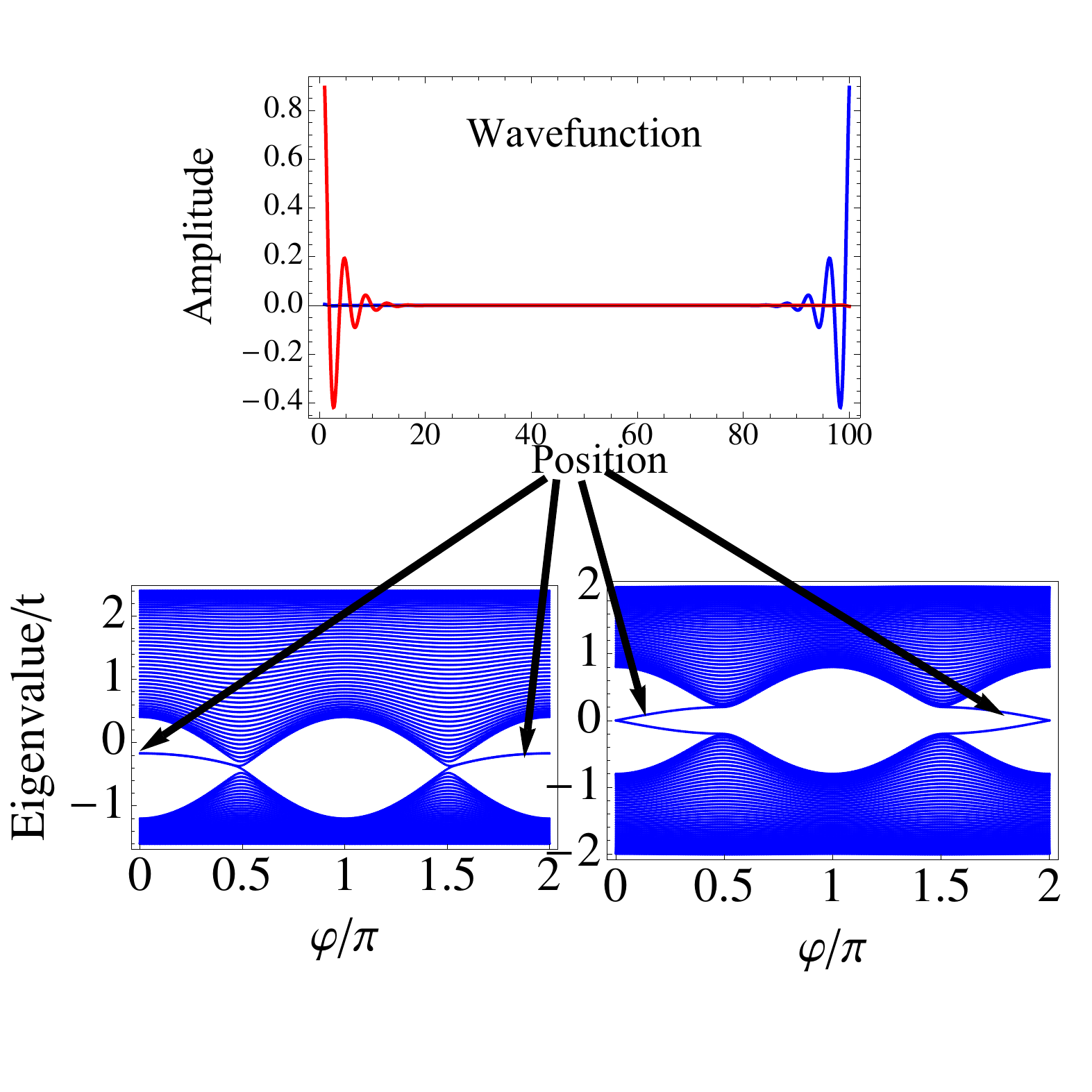}
\vspace{-0.7cm}
\hspace{1cm}\textrm{(a)}\hspace{3cm}\textrm{(b)}
\caption{(a)Energy spectrum with next nearest neighbor hopping for $N=100$ sites ($t=1$, $t'=0.2$, $\lambda=0.4$ and $v=0$). (b)Degeneracy of the zero modes is lifted with weak onsite modulation($t=1$, $v=0.2$ and $\lambda=0.4$ )} \label{robustness}
\end{figure}
\end{center}
%%%%%%%%%%%%%%%%%%%%%%%%%%%%%%%%%%%%%%%%%%%%%%%%%%%%%%%%%%%%%%%%%%%%%%%%%%%%%%%%%%%%%%%%%
For odd number of sites, there exists a single Majorana mode localized on one of the edge sites except for the Dirac points as shown in Fig.~(\ref{odd}). This is an even-odd effect due to the chiral symmetry in the off-diagonal AAH model. 

\textit{Robustness.} The generalized AAH model breaks time reversal symmetry\cite{hatsugai90}. We use this feature to test the robustness of these zero-energy states against time reversal breaking terms. As long as the particle-hole symmetry [$c_n\rightarrow (-1)^n c^\dagger_n$ and $c_n^\dagger \rightarrow (-1)^n c_n$] is preserved~\cite{ryu2002}, the mapping to two decoupled Majorana chains remains valid and thus the edge modes are stable with their energy pinned to zero. However, it is worthwhile to point out that such a symmetry can be broken explicitly by the next-nearest-neighbor hopping and a modulating potential energy, which will couple the two Majorana fermion chains together. This coupling will result in the topological index being ill-defined. To test the fate of the edge modes under such relevant perturbations, we introduce next-nearest-neighbor hopping and a modulating potential energy to the off-diagonal AAH model. For simplicity, we assume that the strength of the next-nearest-neighbor hopping is site independent ($H_{NNN}=t' \sum_n c^\dagger_n c_{n+2}+h.c.$). The modulating potential energy is introduced using the onsite term $v$ in Eq.~\eqref{eq:Hamiltonian} and without any loss of generality we assume $\varphi_v=\varphi_{\lambda}=\varphi$ from hereon. 
As shown in Fig.~ \ref{robustness}, the edge modes are found to be very robust, although the energies are shifted away from zero. This robustness is a direct result of the topological nature of their origin which ensures that turning on a small NNN hopping or some other such perturbation cannot immediately destroy the zero energy modes.
%%%%%%%%%%%%%%%%%%%%%%%%%%%%%%%%%%%%%%%%%%%%%%%%%%%%%%%%%%%%%%%%%%%%%%%%%%%%%%%%%%%%%%%%%%?
\begin{center}
\begin{figure}[htb!]
\textrm{(a)}\hspace{4cm}\textrm{(b)}
\includegraphics[width=4cm,height=2.5cm]{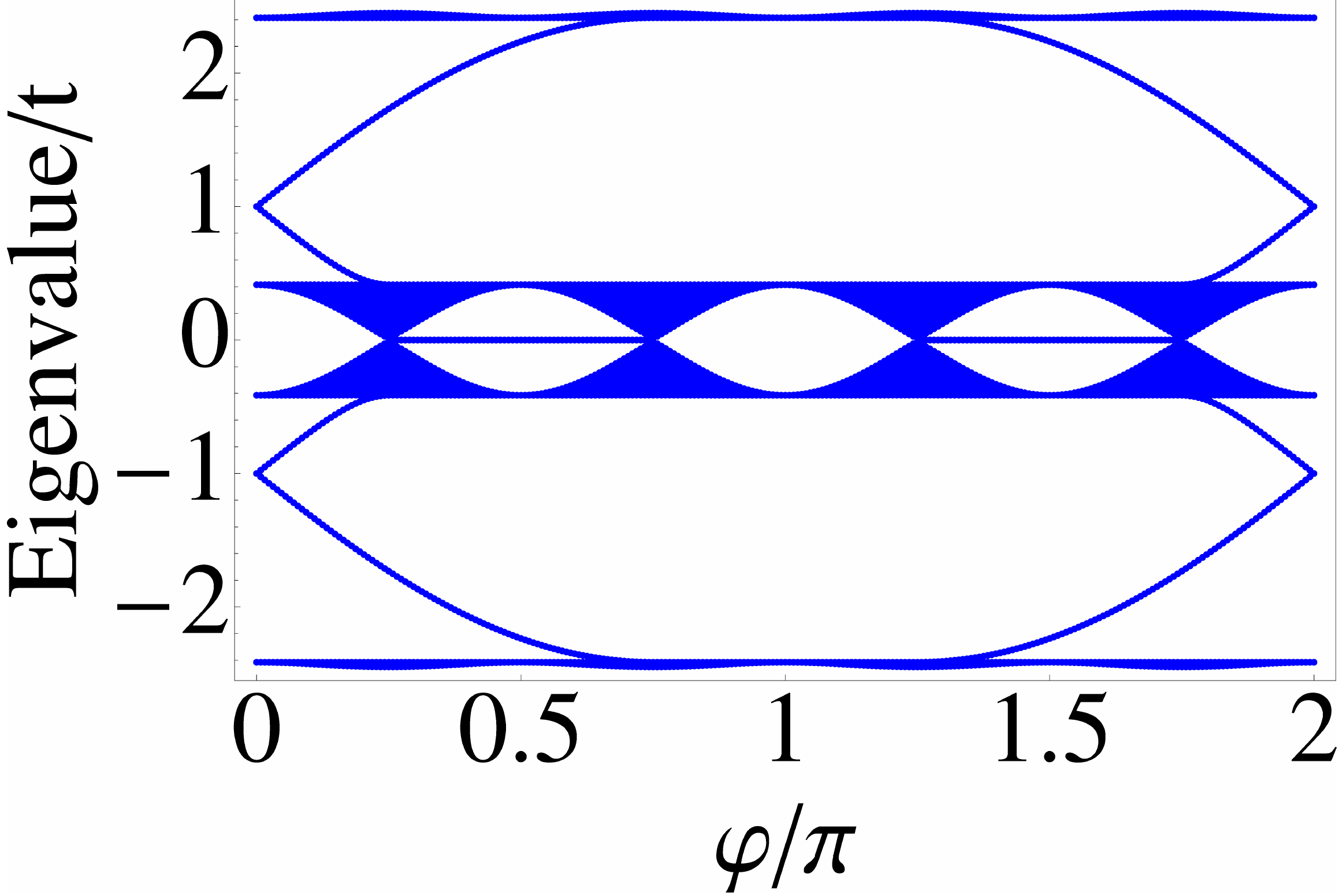}\hspace{0.1cm}\includegraphics[width=4.0cm,height=2.5cm]{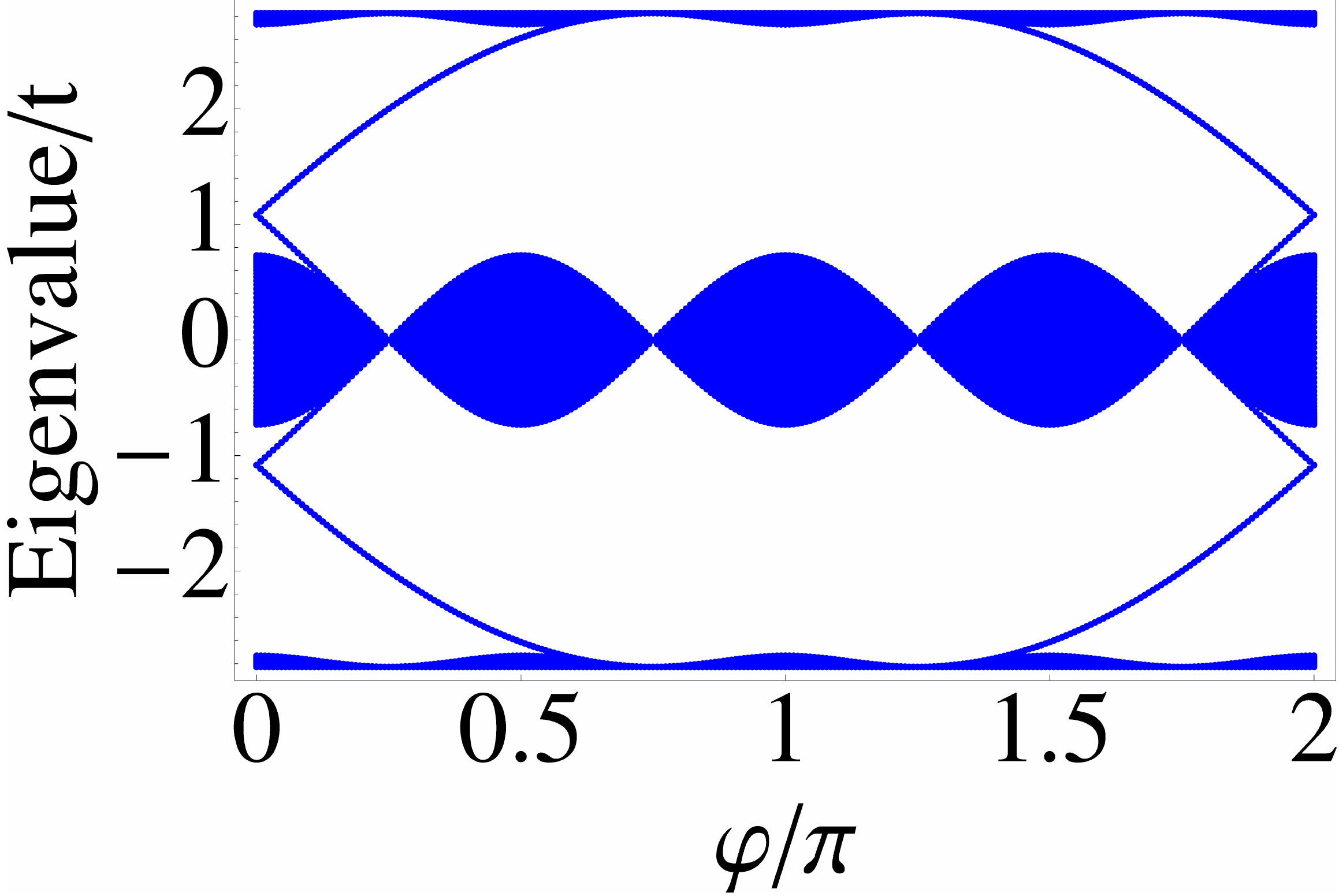}\\
\hspace{-0.2cm}\textrm{(c)}\hspace{4cm}\textrm{(d)}
\includegraphics[width=4.0cm,height=2.5cm]{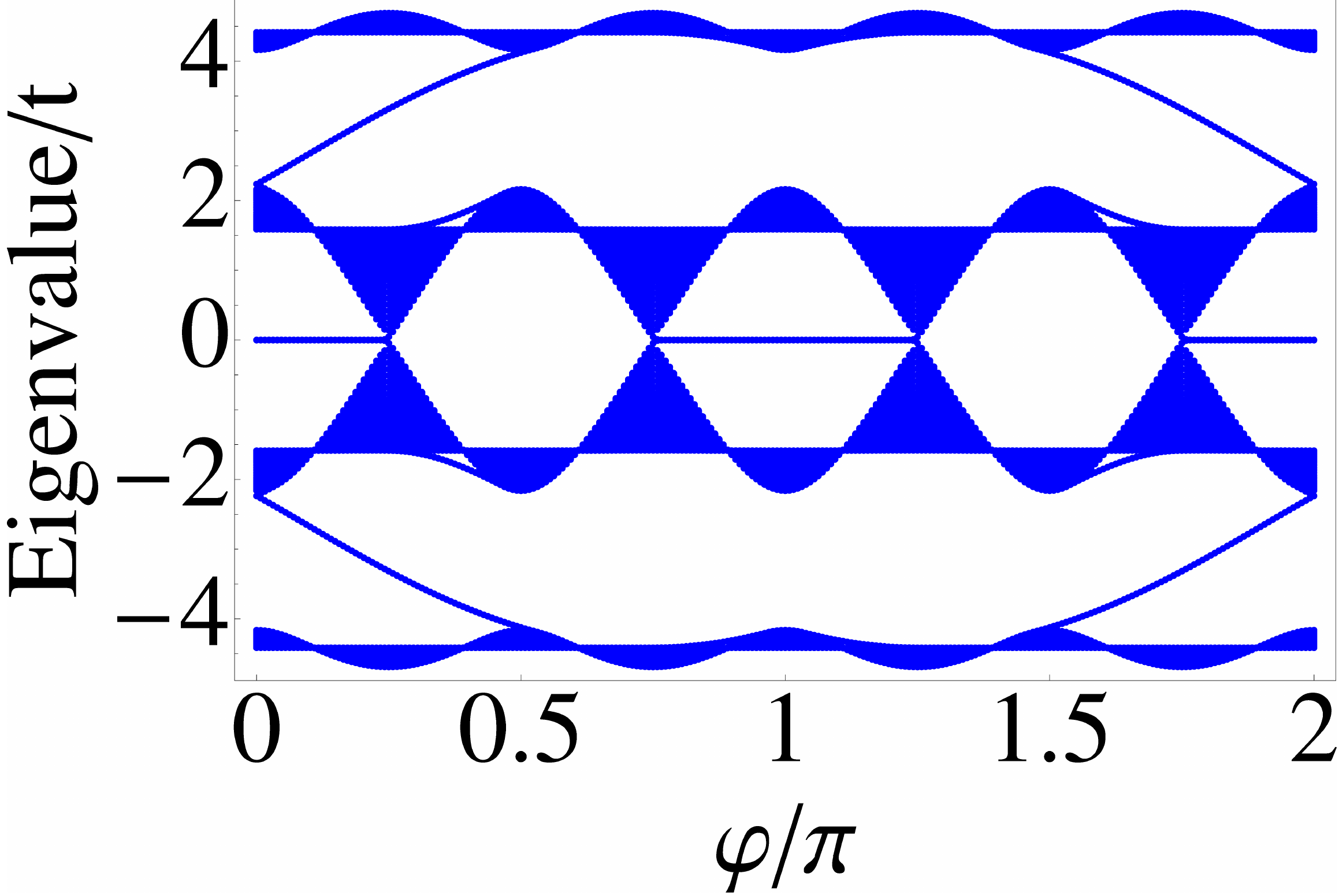}\hspace{0.1cm}\includegraphics[width=4.0cm,height=2.5cm]{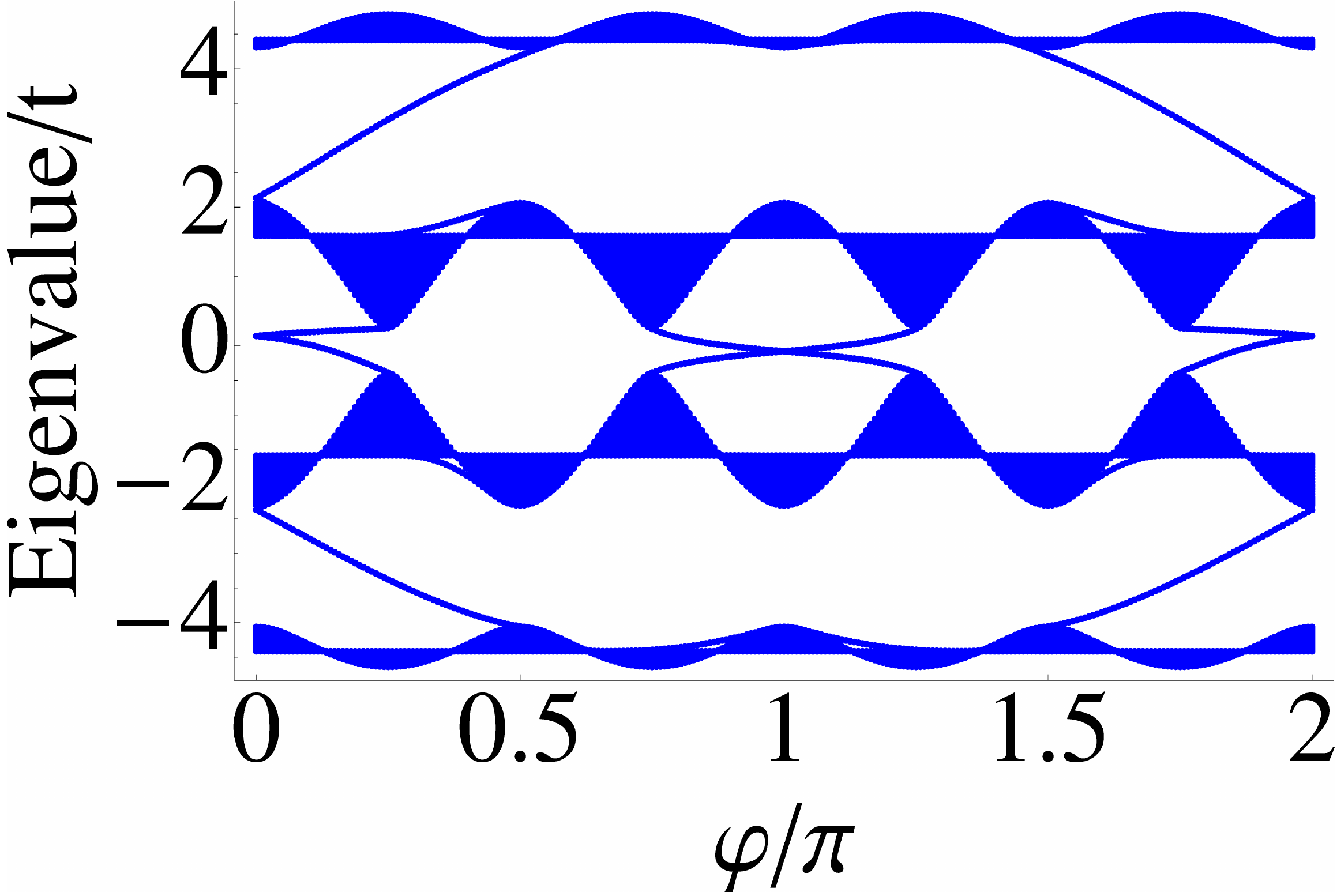}
\caption{Tight binding energy spectrum with open boundary conditions for 100 sites with $b=1/4$ and $t=1$.(a)Energy spectrum as a function of  the parameter $\varphi$ with $\lambda=1.0$ and $v=0$. (b) Central gap closes for  $\lambda=\sqrt{2}$. (c) Energy spectrum with $\lambda=3.0$. Position of degenerate zero modes shifts by $\pi/2$ (d) $t'=0.2$ opens up a gap in the bulk with gapless edge modes of QH type connecting the bulk bands.}
\label{energystates1/4}
\end{figure}
\end{center}
%%%%%%%%%%%%%%%%%%%%%%%%%%%%%%%%%%%%%%%%%%%%%%%%%%%%%%%%%%%%%%%%%%%%%%%%%%%%%%%%%%%%%%%%%
\vspace{-0.3cm}{\it Generic off-diagonal AAH models.} The above analysis can be generalized for the case of $b=1/(2q)$, where $q$ is a positive integer. For example, the off-diagonal AAH model with $b=1/4$ has four energy bands. The top and bottom bands are fully gapped but the two bands in the middle have four band crossing (Dirac) points located at $\varphi=\pi/4$, $3 \pi/4$, $5 \pi/4$ and $7 \pi/4$, as shown in Fig.~\ref{energystates1/4}. Inside the band gap between the top two bands (or between the bottom two bands), the quantum Hall edges states can be observed.
Between the two central bands, zero-energy edge states are observed. For $\lambda/t<\sqrt{2}$, the zero-energy edge modes are found for 
$ \frac{\pi}{4}<\varphi<\frac{3\pi}{4}$ and $ \frac{5\pi}{4}<\varphi<\frac{7\pi}{4}$. When $\lambda/t>\sqrt{2}$, the zero modes are found for 
$-\frac{\pi}{4}<\varphi<\frac{\pi}{4}$ and $ \frac{3\pi}{4}<\varphi<\frac{5\pi}{4}$. The marginal case with $\lambda/t=\sqrt{2}$ shows no gap at any value of $\varphi$ and thus has no edge states between these two bands.
These zero-energy states are of the same origin as the zero-energy edge states discussed above for $b=1/2$. Writing $b=1/4$ case in the Majorana basis  we obtain 
\bea
H&=&\sum_{n}[\Delta_{2n-1}(\gamma_{2n}\gamma_{2n-1})+\Delta_{2n}(\gamma_{2n}\gamma_{2n+1})]\nonumber\\
   &-&\sum_{n}[\Delta_{2n-1}(\tau_{2n}\tau_{2n-1})+\Delta_{2n}(\tau_{2n}\tau_{2n+1})],
               \la{majorana_1/4}
\eea
where $\Delta_{n}=2it[1+\lambda\cos(\varphi+n\frac{\pi}{2})]$.
The zero-energy modes exist for the parameter range satisfying $|1-\lambda\cos\varphi|>|1-\lambda\sin\varphi|$ which is in prefect agreement with the numerical results. 
For $q>1$ the next nearest neighbor hopping term ($t'$) or diagonal  cosine modulation ($v$) opens a gap between the two central bands, and the zero-energy edge modes adiabatically turn into mid-gap edge modes. (Fig.~\ref{energystates1/4}.d). \\
\textit{Experimental realization.} Topological properties of the half flux state can be realized in photonic crystals using setups demonstrated in Ref.~\cite{lahini2009,kraus1} and cold atomic gases using double-well potentials\cite{porto06, porto07,folling07,wirth11}. The details of the experimental realization are shown in the supplementary information~\cite{supp}. In addition, it should also be possible to experimentally study our proposal by using suitably designed  cold atom optical lattices~\cite{billy08,roati08} or semiconductor structures~\cite{sankarprl86,merlin85} where 1D AAH-type quantum systems have earlier been realized in the laboratory.

\textit{Conclusion.} In this work we have unearthed a novel topological aspect of the {\it commensurate off-diagonal} AAH model (i.e. 1D bichromatic lattice model). It is shown both analytically and numerically that the $b=1/(2q)$ flux state of the off-diagonal AAH model supports a topologically non-trivial zero-energy edge modes with respect to the higher dimensional phase parameter $\varphi_{\lambda}$. The topological nature of the commensurate bichromatic 1D Aubry-Andre-Harper model uncovered by us shows some deep and surprising  connections between simple 1D hopping models  and spin liquids, graphene,  1D topological superconductivity. In addition, at $b=1/2$, off-diagonal AAH model can be mapped to the topologically non-trivial polyacetylene (SSH) model\cite{ssh79}. However, such a mapping cannot be generalized to other values of $b$, where also we establish the AAH model to be topological, thus providing a generalization of the SSH model to a new class of topological models deeply connected to the AAH model.\\
\textit{Acknowledgements.} Authors would like to thank Jay D. Sau for insightful comments. This work is supported by JQI-NSF-PFC, ARO-MURI, and AFOSR-MURI.

 \section{Supplementary material for ``Topological zero-energy modes in gapless commensurate Aubry-Andr\'e-Harper models".}
\section{Experimental setup}
\begin{center}
\begin{figure}[htb!]
\includegraphics[width=7.0cm,height=3cm]{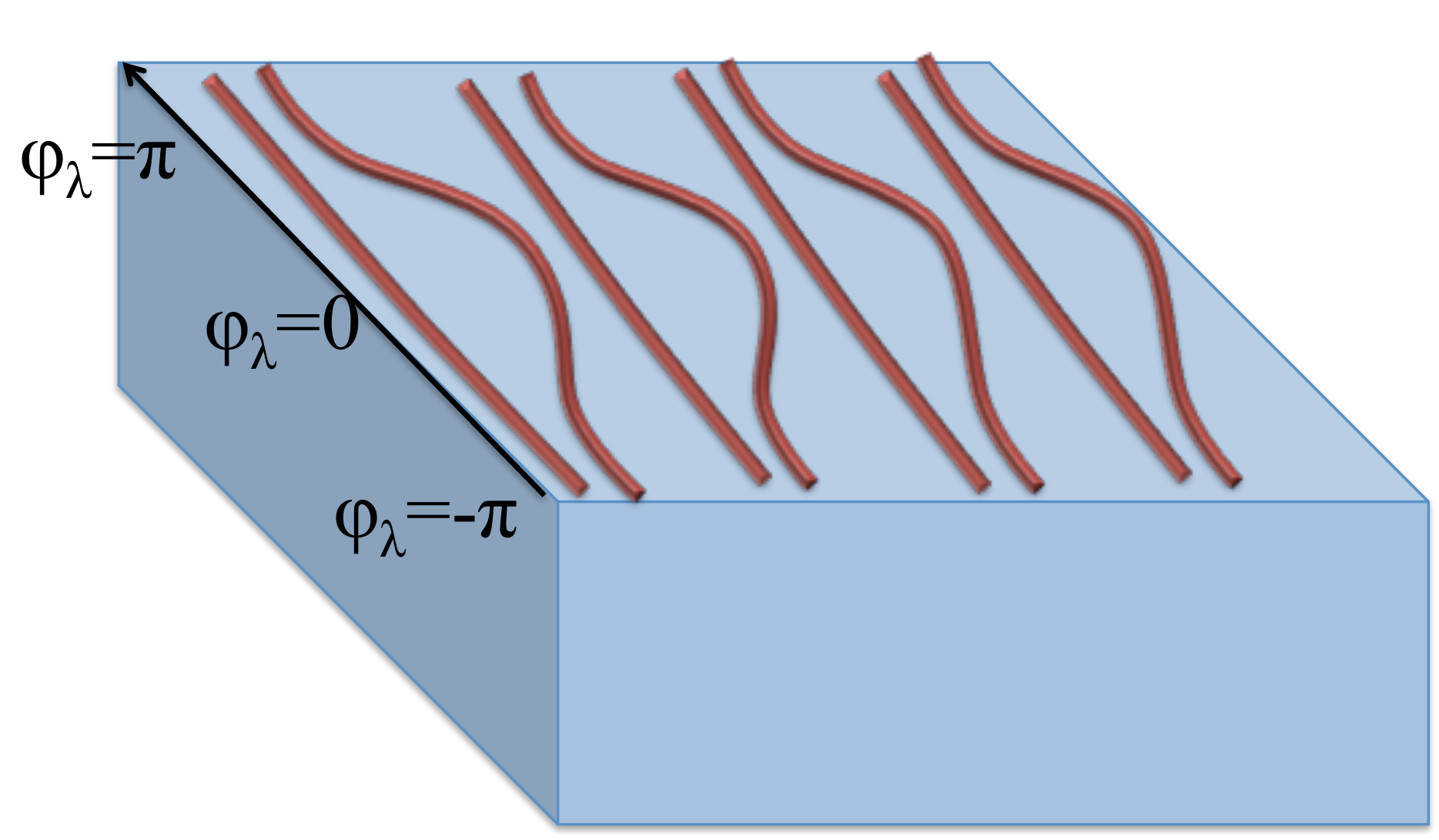}
\caption{Experimental schematic: The red lines shows the single mode waveguides. The hopping amplitude is modulated by varying the spacing between the waveguides.  The phase $\varphi_{\lambda}$ can be scanned along the length of  the waveguide by smoothly changing the spacing between successive channels.}
\label{schematic} 
\end{figure}
\end{center}

In the main paper\cite{vega10,paesani10,briesta10,zheng11}, we discussed the existence of a nontrivial topological phase in the off-diagonal version of 1D Aubre-Andr\'e-Harper (AAH) model.  In this section we present an experimental setup that can realize this non-trivial topology. Fig.~\ref{schematic} shows series of single mode waveguides that represent sites of a tight binding lattice. Each waveguide corresponds to a site in the tight binding Hamiltonian. The potential energy at each site can be adjusted by tuning the width of each waveguide, which offers a way to control $v$ and $\varphi_v$. The off-diagonal modulation $\lambda$ and $\varphi_{\lambda}$ can be controlled by tuning the spacing between two neighboring waveguides, which determines the hopping strengths. By varying these spacings, one can tune the value of $\varphi_{\lambda}$ into the topological regime $-\pi/2<\varphi_{\lambda}<\pi/2$ to observe the topological edge modes discussed in the main text. Experimentally, one has to measure the intensity distributions as a function of the position of the waveguides at different stages of the adiabatic evolution, i.e., different propagation distances. This process in terms of our theoretical analysis amounts to the scanning of the variable $\varphi$. The diagonal parameter is scanned by injecting light into each waveguide one by one and  then monitoring the exiting intensity distributions. The exiting intensity distributions are localized at the end waveguides for topologically non-trivial modulations. In our case the diagonal modulation would simply result in extended intensity distributions. 
%%%%%%%%%%%%%%%%%%%%%%%%%%%%%%%%%%%%%%%%%%%%%%%%%%%%%%%%%%%%%%%%%%%%%%%%%%%%%%%%%%%%%%%%%%?
%%%%%%%%%%%%%%%%%%%%%%%%%%%%%%%%%%%%%%%%%%%%%%%%%%%%%%%%%%%%%%%%%%%%%%%%%%%%%%%%%%%%%%%%%%%

For the case of off-diagonal modulation, the experiment is performed at different slices normal to the $\varphi$ direction. The exiting intensities can be observed only once after the light is injected on one of the edge waveguides via a single laser beam. Three setups with identical spacing modulations are created at different values of $\varphi$. Hence with a single injection on one of the edges the full adiabatic evolution of the light propagation can be mapped out at different values of $\varphi$.\\
In addition to the photonics crystals, this non trivial topology may be observed in cold atomic gases. For example, the $b=1/2$ off-diagonal AAH model can be realized in 1D double well lattice. A bipartite lattice can be achieved by making wells with alternating weak and strong tunneling amplitudes between the successive neighbors. Similar setups can be found in Refs.~(\onlinecite{porto06,porto07,folling07,wirth11}). Generic off-diagonal AAH models can be realized by creating a more complicated pattern of tunneling strengths along the 1D chain. Observing a sharp edge for bosons to establish non-trivial topology is a challenging experimental task, but for fermions, the edge states are located between the lowest and second bands. It should be relatively less complicated to observe the edge state using fermions by filling up the lowest two bands where the proposed non-trivial topology manifests.

%%%%%%%%%%%%%%%%%%
%%%%%%%%%%%%%%%%%
%%%%%%%%%%%%%%%%%%%%%%%%%%%%%%%%%%%%%%%
%%%%%%%%%%%%%%%%%%%%%%%%%%%%%%%%%%%%%%%
%%%%%%%%%%%%%%%%%%%%%%%%%%%%%%%%%%%%%%%
%\begin{thebibliography}{99}

%\bibliographystyle{my-refs}

%\bibliographystyle{plain}
%\bibliographystyle{unsrt}
%\bibliographystyle{iopart-num}
%\bibliographystyle{abbrv}
%\bibliographystyle{apsrev}
%\bibliographystyle{amsplain}
%\bibliographystyle{ieeetr}
%\bibliographystyle{apsrev4-1}
%\bibliographystyle{aipnum4-1.bst}

%\bibliographystyle{unsrtnat}

\bibliography{references.bib}

\begin{thebibliography}{46}%
\makeatletter
\providecommand \@ifxundefined [1]{%
 \@ifx{#1\undefined}
}%
\providecommand \@ifnum [1]{%
 \ifnum #1\expandafter \@firstoftwo
 \else \expandafter \@secondoftwo
 \fi
}%
\providecommand \@ifx [1]{%
 \ifx #1\expandafter \@firstoftwo
 \else \expandafter \@secondoftwo
 \fi
}%
\providecommand \natexlab [1]{#1}%
\providecommand \enquote  [1]{``#1''}%
\providecommand \bibnamefont  [1]{#1}%
\providecommand \bibfnamefont [1]{#1}%
\providecommand \citenamefont [1]{#1}%
\providecommand \href@noop [0]{\@secondoftwo}%
\providecommand \href [0]{\begingroup \@sanitize@url \@href}%
\providecommand \@href[1]{\@@startlink{#1}\@@href}%
\providecommand \@@href[1]{\endgroup#1\@@endlink}%
\providecommand \@sanitize@url [0]{\catcode `\\12\catcode `\$12\catcode
  `\&12\catcode `\#12\catcode `\^12\catcode `\_12\catcode `\%12\relax}%
\providecommand \@@startlink[1]{}%
\providecommand \@@endlink[0]{}%
\providecommand \url  [0]{\begingroup\@sanitize@url \@url }%
\providecommand \@url [1]{\endgroup\@href {#1}{\urlprefix }}%
\providecommand \urlprefix  [0]{URL }%
\providecommand \Eprint [0]{\href }%
\providecommand \doibase [0]{http://dx.doi.org/}%
\providecommand \selectlanguage [0]{\@gobble}%
\providecommand \bibinfo  [0]{\@secondoftwo}%
\providecommand \bibfield  [0]{\@secondoftwo}%
\providecommand \translation [1]{[#1]}%
\providecommand \BibitemOpen [0]{}%
\providecommand \bibitemStop [0]{}%
\providecommand \bibitemNoStop [0]{.\EOS\space}%
\providecommand \EOS [0]{\spacefactor3000\relax}%
\providecommand \BibitemShut  [1]{\csname bibitem#1\endcsname}%
\let\auto@bib@innerbib\@empty
%</preamble>
\bibitem [{\citenamefont {Anderson}(1958)}]{Anderson}%
  \BibitemOpen
  \bibfield  {author} {\bibinfo {author} {\bibfnamefont {P.~W.}\ \bibnamefont
  {Anderson}},\ }\href {\doibase 10.1103/PhysRev.109.1492} {\bibfield
  {journal} {\bibinfo  {journal} {Phys. Rev.}\ }\textbf {\bibinfo {volume}
  {109}},\ \bibinfo {pages} {1492} (\bibinfo {year} {1958})}\BibitemShut
  {NoStop}%
\bibitem [{\citenamefont {Aubry}\ and\ \citenamefont {Andr\'e}(1980)}]{AA}%
  \BibitemOpen
  \bibfield  {author} {\bibinfo {author} {\bibfnamefont {S.}~\bibnamefont
  {Aubry}}\ and\ \bibinfo {author} {\bibfnamefont {G.}~\bibnamefont
  {Andr\'e}},\ }\href@noop {} {\bibfield  {journal} {\bibinfo  {journal} {Ann.
  Isr. Phys. Soc.}\ }\textbf {\bibinfo {volume} {3}} (\bibinfo {year}
  {1980})}\BibitemShut {NoStop}%
\bibitem [{\citenamefont {Harper}(1955)}]{harper55}%
  \BibitemOpen
  \bibfield  {author} {\bibinfo {author} {\bibfnamefont {P.~G.}\ \bibnamefont
  {Harper}},\ }\href {http://stacks.iop.org/0370-1298/68/i=10/a=304} {\bibfield
   {journal} {\bibinfo  {journal} {Proceedings of the Physical Society. Section
  A}\ }\textbf {\bibinfo {volume} {68}},\ \bibinfo {pages} {874} (\bibinfo
  {year} {1955})}\BibitemShut {NoStop}%
\bibitem [{\citenamefont {Das~Sarma}\ \emph {et~al.}(1988)\citenamefont
  {Das~Sarma}, \citenamefont {He},\ and\ \citenamefont {Xie}}]{sankarprl88}%
  \BibitemOpen
  \bibfield  {author} {\bibinfo {author} {\bibfnamefont {S.}~\bibnamefont
  {Das~Sarma}}, \bibinfo {author} {\bibfnamefont {S.}~\bibnamefont {He}}, \
  and\ \bibinfo {author} {\bibfnamefont {X.~C.}\ \bibnamefont {Xie}},\ }\href
  {\doibase 10.1103/PhysRevLett.61.2144} {\bibfield  {journal} {\bibinfo
  {journal} {Phys. Rev. Lett.}\ }\textbf {\bibinfo {volume} {61}},\ \bibinfo
  {pages} {2144} (\bibinfo {year} {1988})}\BibitemShut {NoStop}%
\bibitem [{\citenamefont {Thouless}(1988)}]{thoulessprl88}%
  \BibitemOpen
  \bibfield  {author} {\bibinfo {author} {\bibfnamefont {D.~J.}\ \bibnamefont
  {Thouless}},\ }\href {\doibase 10.1103/PhysRevLett.61.2141} {\bibfield
  {journal} {\bibinfo  {journal} {Phys. Rev. Lett.}\ }\textbf {\bibinfo
  {volume} {61}},\ \bibinfo {pages} {2141} (\bibinfo {year}
  {1988})}\BibitemShut {NoStop}%
\bibitem [{\citenamefont {Das~Sarma}\ \emph {et~al.}(1990)\citenamefont
  {Das~Sarma}, \citenamefont {He},\ and\ \citenamefont {Xie}}]{sankarprb90}%
  \BibitemOpen
  \bibfield  {author} {\bibinfo {author} {\bibfnamefont {S.}~\bibnamefont
  {Das~Sarma}}, \bibinfo {author} {\bibfnamefont {S.}~\bibnamefont {He}}, \
  and\ \bibinfo {author} {\bibfnamefont {X.~C.}\ \bibnamefont {Xie}},\ }\href
  {\doibase 10.1103/PhysRevB.41.5544} {\bibfield  {journal} {\bibinfo
  {journal} {Phys. Rev. B}\ }\textbf {\bibinfo {volume} {41}},\ \bibinfo
  {pages} {5544} (\bibinfo {year} {1990})}\BibitemShut {NoStop}%
\bibitem [{\citenamefont {Biddle}\ \emph {et~al.}(2009)\citenamefont {Biddle},
  \citenamefont {Wang}, \citenamefont {Priour},\ and\ \citenamefont
  {Das~Sarma}}]{biddlepra09}%
  \BibitemOpen
  \bibfield  {author} {\bibinfo {author} {\bibfnamefont {J.}~\bibnamefont
  {Biddle}}, \bibinfo {author} {\bibfnamefont {B.}~\bibnamefont {Wang}},
  \bibinfo {author} {\bibfnamefont {D.~J.}\ \bibnamefont {Priour}}, \ and\
  \bibinfo {author} {\bibfnamefont {S.}~\bibnamefont {Das~Sarma}},\ }\href
  {\doibase 10.1103/PhysRevA.80.021603} {\bibfield  {journal} {\bibinfo
  {journal} {Phys. Rev. A}\ }\textbf {\bibinfo {volume} {80}},\ \bibinfo
  {pages} {021603} (\bibinfo {year} {2009})}\BibitemShut {NoStop}%
\bibitem [{\citenamefont {Biddle}\ and\ \citenamefont
  {Das~Sarma}(2010)}]{biddleprl10}%
  \BibitemOpen
  \bibfield  {author} {\bibinfo {author} {\bibfnamefont {J.}~\bibnamefont
  {Biddle}}\ and\ \bibinfo {author} {\bibfnamefont {S.}~\bibnamefont
  {Das~Sarma}},\ }\href {\doibase 10.1103/PhysRevLett.104.070601} {\bibfield
  {journal} {\bibinfo  {journal} {Phys. Rev. Lett.}\ }\textbf {\bibinfo
  {volume} {104}},\ \bibinfo {pages} {070601} (\bibinfo {year}
  {2010})}\BibitemShut {NoStop}%
\bibitem [{\citenamefont {Lahini}\ \emph {et~al.}(2009)\citenamefont {Lahini},
  \citenamefont {Pugatch}, \citenamefont {Pozzi}, \citenamefont {Sorel},
  \citenamefont {Morandotti}, \citenamefont {Davidson},\ and\ \citenamefont
  {Silberberg}}]{lahini2009}%
  \BibitemOpen
  \bibfield  {author} {\bibinfo {author} {\bibfnamefont {Y.}~\bibnamefont
  {Lahini}}, \bibinfo {author} {\bibfnamefont {R.}~\bibnamefont {Pugatch}},
  \bibinfo {author} {\bibfnamefont {F.}~\bibnamefont {Pozzi}}, \bibinfo
  {author} {\bibfnamefont {M.}~\bibnamefont {Sorel}}, \bibinfo {author}
  {\bibfnamefont {R.}~\bibnamefont {Morandotti}}, \bibinfo {author}
  {\bibfnamefont {N.}~\bibnamefont {Davidson}}, \ and\ \bibinfo {author}
  {\bibfnamefont {Y.}~\bibnamefont {Silberberg}},\ }\href {\doibase
  10.1103/PhysRevLett.103.013901} {\bibfield  {journal} {\bibinfo  {journal}
  {Phys. Rev. Lett.}\ }\textbf {\bibinfo {volume} {103}},\ \bibinfo {pages}
  {013901} (\bibinfo {year} {2009})}\BibitemShut {NoStop}%
\bibitem [{\citenamefont {Kraus}\ \emph {et~al.}(2012)\citenamefont {Kraus},
  \citenamefont {Lahini}, \citenamefont {Ringel}, \citenamefont {Verbin},\ and\
  \citenamefont {Zilberberg}}]{kraus1}%
  \BibitemOpen
  \bibfield  {author} {\bibinfo {author} {\bibfnamefont {Y.~E.}\ \bibnamefont
  {Kraus}}, \bibinfo {author} {\bibfnamefont {Y.}~\bibnamefont {Lahini}},
  \bibinfo {author} {\bibfnamefont {Z.}~\bibnamefont {Ringel}}, \bibinfo
  {author} {\bibfnamefont {M.}~\bibnamefont {Verbin}}, \ and\ \bibinfo {author}
  {\bibfnamefont {O.}~\bibnamefont {Zilberberg}},\ }\href {\doibase
  10.1103/PhysRevLett.109.106402} {\bibfield  {journal} {\bibinfo  {journal}
  {Phys. Rev. Lett.}\ }\textbf {\bibinfo {volume} {109}},\ \bibinfo {pages}
  {106402} (\bibinfo {year} {2012})}\BibitemShut {NoStop}%
\bibitem [{\citenamefont {Hofstadter}(1976)}]{hofstadter}%
  \BibitemOpen
  \bibfield  {author} {\bibinfo {author} {\bibfnamefont {D.~R.}\ \bibnamefont
  {Hofstadter}},\ }\href {\doibase 10.1103/PhysRevB.14.2239} {\bibfield
  {journal} {\bibinfo  {journal} {Phys. Rev. B}\ }\textbf {\bibinfo {volume}
  {14}},\ \bibinfo {pages} {2239} (\bibinfo {year} {1976})}\BibitemShut
  {NoStop}%
\bibitem [{\citenamefont {Laughlin}(1981)}]{laughlin}%
  \BibitemOpen
  \bibfield  {author} {\bibinfo {author} {\bibfnamefont {R.~B.}\ \bibnamefont
  {Laughlin}},\ }\href {\doibase 10.1103/PhysRevB.23.5632} {\bibfield
  {journal} {\bibinfo  {journal} {Phys. Rev. B}\ }\textbf {\bibinfo {volume}
  {23}},\ \bibinfo {pages} {5632} (\bibinfo {year} {1981})}\BibitemShut
  {NoStop}%
\bibitem [{\citenamefont {Thouless}\ \emph {et~al.}(1982)\citenamefont
  {Thouless}, \citenamefont {Kohmoto}, \citenamefont {Nightingale},\ and\
  \citenamefont {den Nijs}}]{tknn}%
  \BibitemOpen
  \bibfield  {author} {\bibinfo {author} {\bibfnamefont {D.~J.}\ \bibnamefont
  {Thouless}}, \bibinfo {author} {\bibfnamefont {M.}~\bibnamefont {Kohmoto}},
  \bibinfo {author} {\bibfnamefont {M.~P.}\ \bibnamefont {Nightingale}}, \ and\
  \bibinfo {author} {\bibfnamefont {M.}~\bibnamefont {den Nijs}},\ }\href
  {\doibase 10.1103/PhysRevLett.49.405} {\bibfield  {journal} {\bibinfo
  {journal} {Phys. Rev. Lett.}\ }\textbf {\bibinfo {volume} {49}},\ \bibinfo
  {pages} {405} (\bibinfo {year} {1982})}\BibitemShut {NoStop}%
\bibitem [{\citenamefont {Avron}\ \emph {et~al.}(1986)\citenamefont {Avron},
  \citenamefont {Seiler},\ and\ \citenamefont {Shapiro}}]{Avron}%
  \BibitemOpen
  \bibfield  {author} {\bibinfo {author} {\bibfnamefont {Y.}~\bibnamefont
  {Avron}}, \bibinfo {author} {\bibfnamefont {R.}~\bibnamefont {Seiler}}, \
  and\ \bibinfo {author} {\bibfnamefont {B.}~\bibnamefont {Shapiro}},\ }\href
  {\doibase 10.1016/0550-3213(86)90315-9} {\bibfield  {journal} {\bibinfo
  {journal} {Nuclear Physics B}\ }\textbf {\bibinfo {volume} {265}},\ \bibinfo
  {pages} {364 } (\bibinfo {year} {1986})}\BibitemShut {NoStop}%
\bibitem [{\citenamefont {Hatsugai}(1993)}]{hatsugai}%
  \BibitemOpen
  \bibfield  {author} {\bibinfo {author} {\bibfnamefont {Y.}~\bibnamefont
  {Hatsugai}},\ }\href {\doibase 10.1103/PhysRevLett.71.3697} {\bibfield
  {journal} {\bibinfo  {journal} {Phys. Rev. Lett.}\ }\textbf {\bibinfo
  {volume} {71}},\ \bibinfo {pages} {3697} (\bibinfo {year}
  {1993})}\BibitemShut {NoStop}%
\bibitem [{\citenamefont {Kraus}\ and\ \citenamefont
  {Zilberberg}(2012)}]{kraus2}%
  \BibitemOpen
  \bibfield  {author} {\bibinfo {author} {\bibfnamefont {Y.~E.}\ \bibnamefont
  {Kraus}}\ and\ \bibinfo {author} {\bibfnamefont {O.}~\bibnamefont
  {Zilberberg}},\ }\href {\doibase 10.1103/PhysRevLett.109.116404} {\bibfield
  {journal} {\bibinfo  {journal} {Phys. Rev. Lett.}\ }\textbf {\bibinfo
  {volume} {109}},\ \bibinfo {pages} {116404} (\bibinfo {year}
  {2012})}\BibitemShut {NoStop}%
\bibitem [{\citenamefont {Lang}\ \emph {et~al.}(2012)\citenamefont {Lang},
  \citenamefont {Cai},\ and\ \citenamefont {Chen}}]{langprl12}%
  \BibitemOpen
  \bibfield  {author} {\bibinfo {author} {\bibfnamefont {L.-J.}\ \bibnamefont
  {Lang}}, \bibinfo {author} {\bibfnamefont {X.}~\bibnamefont {Cai}}, \ and\
  \bibinfo {author} {\bibfnamefont {S.}~\bibnamefont {Chen}},\ }\href {\doibase
  10.1103/PhysRevLett.108.220401} {\bibfield  {journal} {\bibinfo  {journal}
  {Phys. Rev. Lett.}\ }\textbf {\bibinfo {volume} {108}},\ \bibinfo {pages}
  {220401} (\bibinfo {year} {2012})}\BibitemShut {NoStop}%
\bibitem [{\citenamefont {Lang}\ and\ \citenamefont {Chen}(2012)}]{lang12}%
  \BibitemOpen
  \bibfield  {author} {\bibinfo {author} {\bibfnamefont {L.-J.}\ \bibnamefont
  {Lang}}\ and\ \bibinfo {author} {\bibfnamefont {S.}~\bibnamefont {Chen}},\
  }\href {\doibase 10.1103/PhysRevB.86.205135} {\bibfield  {journal} {\bibinfo
  {journal} {Phys. Rev. B}\ }\textbf {\bibinfo {volume} {86}},\ \bibinfo
  {pages} {205135} (\bibinfo {year} {2012})}\BibitemShut {NoStop}%
\bibitem [{\citenamefont {Han}\ \emph {et~al.}(1994)\citenamefont {Han},
  \citenamefont {Thouless}, \citenamefont {Hiramoto},\ and\ \citenamefont
  {Kohmoto}}]{han94}%
  \BibitemOpen
  \bibfield  {author} {\bibinfo {author} {\bibfnamefont {J.~H.}\ \bibnamefont
  {Han}}, \bibinfo {author} {\bibfnamefont {D.~J.}\ \bibnamefont {Thouless}},
  \bibinfo {author} {\bibfnamefont {H.}~\bibnamefont {Hiramoto}}, \ and\
  \bibinfo {author} {\bibfnamefont {M.}~\bibnamefont {Kohmoto}},\ }\href
  {\doibase 10.1103/PhysRevB.50.11365} {\bibfield  {journal} {\bibinfo
  {journal} {Phys. Rev. B}\ }\textbf {\bibinfo {volume} {50}},\ \bibinfo
  {pages} {11365} (\bibinfo {year} {1994})}\BibitemShut {NoStop}%
\bibitem [{\citenamefont {Ryu}\ and\ \citenamefont {Hatsugai}(2002)}]{ryu2002}%
  \BibitemOpen
  \bibfield  {author} {\bibinfo {author} {\bibfnamefont {S.}~\bibnamefont
  {Ryu}}\ and\ \bibinfo {author} {\bibfnamefont {Y.}~\bibnamefont {Hatsugai}},\
  }\href {\doibase 10.1103/PhysRevLett.89.077002} {\bibfield  {journal}
  {\bibinfo  {journal} {Phys. Rev. Lett.}\ }\textbf {\bibinfo {volume} {89}},\
  \bibinfo {pages} {077002} (\bibinfo {year} {2002})}\BibitemShut {NoStop}%
\bibitem [{\citenamefont {Esaki}\ \emph {et~al.}(2009)\citenamefont {Esaki},
  \citenamefont {Sato}, \citenamefont {Kohmoto},\ and\ \citenamefont
  {Halperin}}]{halperin09}%
  \BibitemOpen
  \bibfield  {author} {\bibinfo {author} {\bibfnamefont {K.}~\bibnamefont
  {Esaki}}, \bibinfo {author} {\bibfnamefont {M.}~\bibnamefont {Sato}},
  \bibinfo {author} {\bibfnamefont {M.}~\bibnamefont {Kohmoto}}, \ and\
  \bibinfo {author} {\bibfnamefont {B.~I.}\ \bibnamefont {Halperin}},\ }\href
  {\doibase 10.1103/PhysRevB.80.125405} {\bibfield  {journal} {\bibinfo
  {journal} {Phys. Rev. B}\ }\textbf {\bibinfo {volume} {80}},\ \bibinfo
  {pages} {125405} (\bibinfo {year} {2009})}\BibitemShut {NoStop}%
\bibitem [{\citenamefont {Delplace}\ and\ \citenamefont
  {Montambaux}(2010)}]{montambaux10}%
  \BibitemOpen
  \bibfield  {author} {\bibinfo {author} {\bibfnamefont {P.}~\bibnamefont
  {Delplace}}\ and\ \bibinfo {author} {\bibfnamefont {G.}~\bibnamefont
  {Montambaux}},\ }\href {\doibase 10.1103/PhysRevB.82.035438} {\bibfield
  {journal} {\bibinfo  {journal} {Phys. Rev. B}\ }\textbf {\bibinfo {volume}
  {82}},\ \bibinfo {pages} {035438} (\bibinfo {year} {2010})}\BibitemShut
  {NoStop}%
\bibitem [{\citenamefont {Kitaev}(2001)}]{kitaev01}%
  \BibitemOpen
  \bibfield  {author} {\bibinfo {author} {\bibfnamefont {A.~Y.}\ \bibnamefont
  {Kitaev}},\ }\href@noop {} {\bibfield  {journal} {\bibinfo  {journal}
  {Phys.-Usp.}\ }\textbf {\bibinfo {volume} {44}} (\bibinfo {year}
  {2001})}\BibitemShut {NoStop}%
\bibitem [{\citenamefont {Su}\ \emph {et~al.}(1979)\citenamefont {Su},
  \citenamefont {Schrieffer},\ and\ \citenamefont {Heeger}}]{ssh79}%
  \BibitemOpen
  \bibfield  {author} {\bibinfo {author} {\bibfnamefont {W.~P.}\ \bibnamefont
  {Su}}, \bibinfo {author} {\bibfnamefont {J.~R.}\ \bibnamefont {Schrieffer}},
  \ and\ \bibinfo {author} {\bibfnamefont {A.~J.}\ \bibnamefont {Heeger}},\
  }\href {\doibase 10.1103/PhysRevLett.42.1698} {\bibfield  {journal} {\bibinfo
   {journal} {Phys. Rev. Lett.}\ }\textbf {\bibinfo {volume} {42}},\ \bibinfo
  {pages} {1698} (\bibinfo {year} {1979})}\BibitemShut {NoStop}%
\bibitem [{\citenamefont {Satija}\ and\ \citenamefont
  {Naumis}(2012)}]{diptiman12}%
  \BibitemOpen
  \bibfield  {author} {\bibinfo {author} {\bibfnamefont {I.~I.}\ \bibnamefont
  {Satija}}\ and\ \bibinfo {author} {\bibfnamefont {G.~G.}\ \bibnamefont
  {Naumis}},\ }\href@noop {} {\bibfield  {journal} {\bibinfo  {journal}
  {arXiv}\ }\textbf {\bibinfo {volume} {1210.5159}} (\bibinfo {year}
  {2012})}\BibitemShut {NoStop}%
\bibitem [{\citenamefont {DeGottardi}\ \emph {et~al.}(2012)\citenamefont
  {DeGottardi}, \citenamefont {Sen},\ and\ \citenamefont
  {Vishveshwara}}]{indu12}%
  \BibitemOpen
  \bibfield  {author} {\bibinfo {author} {\bibfnamefont {W.}~\bibnamefont
  {DeGottardi}}, \bibinfo {author} {\bibfnamefont {D.}~\bibnamefont {Sen}}, \
  and\ \bibinfo {author} {\bibfnamefont {S.}~\bibnamefont {Vishveshwara}},\
  }\href@noop {} {\bibfield  {journal} {\bibinfo  {journal} {arXiv}\ }\textbf
  {\bibinfo {volume} {1208.0015}} (\bibinfo {year} {2012})}\BibitemShut
  {NoStop}%
\bibitem [{\citenamefont {Hiramoto}\ and\ \citenamefont
  {Kohmoto}(1989)}]{hiramoto89}%
  \BibitemOpen
  \bibfield  {author} {\bibinfo {author} {\bibfnamefont {H.}~\bibnamefont
  {Hiramoto}}\ and\ \bibinfo {author} {\bibfnamefont {M.}~\bibnamefont
  {Kohmoto}},\ }\href {\doibase 10.1103/PhysRevB.40.8225} {\bibfield  {journal}
  {\bibinfo  {journal} {Phys. Rev. B}\ }\textbf {\bibinfo {volume} {40}},\
  \bibinfo {pages} {8225} (\bibinfo {year} {1989})}\BibitemShut {NoStop}%
\bibitem [{\citenamefont {Kohmoto}(1983)}]{kohmoto83}%
  \BibitemOpen
  \bibfield  {author} {\bibinfo {author} {\bibfnamefont {M.}~\bibnamefont
  {Kohmoto}},\ }\href {\doibase 10.1103/PhysRevLett.51.1198} {\bibfield
  {journal} {\bibinfo  {journal} {Phys. Rev. Lett.}\ }\textbf {\bibinfo
  {volume} {51}},\ \bibinfo {pages} {1198} (\bibinfo {year}
  {1983})}\BibitemShut {NoStop}%
\bibitem [{\citenamefont {Wen}(2004)}]{wenbook}%
  \BibitemOpen
  \bibfield  {author} {\bibinfo {author} {\bibfnamefont {X.-G.}\ \bibnamefont
  {Wen}},\ }\href@noop {} {\emph {\bibinfo {title} {Quantum field theory of
  many-body systems: from the origin of sound to an origin of light and
  electrons.}}}\ (\bibinfo  {publisher} {Oxford University Press},\ \bibinfo
  {year} {2004})\BibitemShut {NoStop}%
\bibitem [{\citenamefont {Fradkin}(2013)}]{fradkinbook}%
  \BibitemOpen
  \bibfield  {author} {\bibinfo {author} {\bibfnamefont {E.}~\bibnamefont
  {Fradkin}},\ }\href@noop {} {\emph {\bibinfo {title} {Field Theories of
  Condensed Matter Physics}}},\ Field Theories of Condensed Matter Systems\
  (\bibinfo  {publisher} {Cambridge University Press},\ \bibinfo {year}
  {2013})\BibitemShut {NoStop}%
\bibitem [{\citenamefont {Kitaev}(2009)}]{kitaev09}%
  \BibitemOpen
  \bibfield  {author} {\bibinfo {author} {\bibfnamefont {A.~Y.}\ \bibnamefont
  {Kitaev}},\ }\href@noop {} {\bibfield  {journal} {\bibinfo  {journal} {AIP
  Conf. Proc.}\ }\textbf {\bibinfo {volume} {1134}} (\bibinfo {year}
  {2009})}\BibitemShut {NoStop}%
\bibitem [{\citenamefont {Schnyder}\ \emph {et~al.}(2009)\citenamefont
  {Schnyder}, \citenamefont {Ryu}, \citenamefont {Furusaki},\ and\
  \citenamefont {Ludwig}}]{ryu09}%
  \BibitemOpen
  \bibfield  {author} {\bibinfo {author} {\bibfnamefont {A.~P.}\ \bibnamefont
  {Schnyder}}, \bibinfo {author} {\bibfnamefont {S.}~\bibnamefont {Ryu}},
  \bibinfo {author} {\bibfnamefont {A.}~\bibnamefont {Furusaki}}, \ and\
  \bibinfo {author} {\bibfnamefont {A.~W.~W.}\ \bibnamefont {Ludwig}},\
  }\href@noop {} {\bibfield  {journal} {\bibinfo  {journal} {AIP Conf. Proc.}\
  }\textbf {\bibinfo {volume} {1134}} (\bibinfo {year} {2009})}\BibitemShut
  {NoStop}%
\bibitem [{\citenamefont {Hatsugai}\ and\ \citenamefont
  {Kohmoto}(1990)}]{hatsugai90}%
  \BibitemOpen
  \bibfield  {author} {\bibinfo {author} {\bibfnamefont {Y.}~\bibnamefont
  {Hatsugai}}\ and\ \bibinfo {author} {\bibfnamefont {M.}~\bibnamefont
  {Kohmoto}},\ }\href {\doibase 10.1103/PhysRevB.42.8282} {\bibfield  {journal}
  {\bibinfo  {journal} {Phys. Rev. B}\ }\textbf {\bibinfo {volume} {42}},\
  \bibinfo {pages} {8282} (\bibinfo {year} {1990})}\BibitemShut {NoStop}%
\bibitem [{\citenamefont {Sebby-Strabley}\ \emph {et~al.}(2006)\citenamefont
  {Sebby-Strabley}, \citenamefont {Anderlini}, \citenamefont {Jessen},\ and\
  \citenamefont {Porto}}]{porto06}%
  \BibitemOpen
  \bibfield  {author} {\bibinfo {author} {\bibfnamefont {J.}~\bibnamefont
  {Sebby-Strabley}}, \bibinfo {author} {\bibfnamefont {M.}~\bibnamefont
  {Anderlini}}, \bibinfo {author} {\bibfnamefont {P.~S.}\ \bibnamefont
  {Jessen}}, \ and\ \bibinfo {author} {\bibfnamefont {J.~V.}\ \bibnamefont
  {Porto}},\ }\href {\doibase 10.1103/PhysRevA.73.033605} {\bibfield  {journal}
  {\bibinfo  {journal} {Phys. Rev. A}\ }\textbf {\bibinfo {volume} {73}},\
  \bibinfo {pages} {033605} (\bibinfo {year} {2006})}\BibitemShut {NoStop}%
\bibitem [{\citenamefont {Lee}\ \emph {et~al.}(2007)\citenamefont {Lee},
  \citenamefont {Anderlini}, \citenamefont {Brown}, \citenamefont
  {Sebby-Strabley}, \citenamefont {Phillips},\ and\ \citenamefont
  {Porto}}]{porto07}%
  \BibitemOpen
  \bibfield  {author} {\bibinfo {author} {\bibfnamefont {P.~J.}\ \bibnamefont
  {Lee}}, \bibinfo {author} {\bibfnamefont {M.}~\bibnamefont {Anderlini}},
  \bibinfo {author} {\bibfnamefont {B.~L.}\ \bibnamefont {Brown}}, \bibinfo
  {author} {\bibfnamefont {J.}~\bibnamefont {Sebby-Strabley}}, \bibinfo
  {author} {\bibfnamefont {W.~D.}\ \bibnamefont {Phillips}}, \ and\ \bibinfo
  {author} {\bibfnamefont {J.~V.}\ \bibnamefont {Porto}},\ }\href {\doibase
  10.1103/PhysRevLett.99.020402} {\bibfield  {journal} {\bibinfo  {journal}
  {Phys. Rev. Lett.}\ }\textbf {\bibinfo {volume} {99}},\ \bibinfo {pages}
  {020402} (\bibinfo {year} {2007})}\BibitemShut {NoStop}%
\bibitem [{\citenamefont {F\"oiling~et al.}(2007)}]{folling07}%
  \BibitemOpen
  \bibfield  {author} {\bibinfo {author} {\bibfnamefont {S.}~\bibnamefont
  {F\"oiling~et al.}},\ }\href@noop {} {\bibfield  {journal} {\bibinfo
  {journal} {Nature}\ }\textbf {\bibinfo {volume} {7}},\ \bibinfo {pages} {147}
  (\bibinfo {year} {2007})}\BibitemShut {NoStop}%
\bibitem [{\citenamefont {Wirth~et al.}(2011)}]{wirth11}%
  \BibitemOpen
  \bibfield  {author} {\bibinfo {author} {\bibfnamefont {G.}~\bibnamefont
  {Wirth~et al.}},\ }\href@noop {} {\bibfield  {journal} {\bibinfo  {journal}
  {Nature Physics}\ }\textbf {\bibinfo {volume} {448}},\ \bibinfo {pages}
  {1029} (\bibinfo {year} {2011})}\BibitemShut {NoStop}%
\bibitem [{sup()}]{supp}%
  \BibitemOpen
  \href@noop {} {\bibinfo  {journal} {For more details see supplemental
  information}\ }\BibitemShut {NoStop}%
\bibitem [{\citenamefont {Billy~et al.}(2008)}]{billy08}%
  \BibitemOpen
\bibfield  {journal} {  }\bibfield  {author} {\bibinfo {author} {\bibfnamefont
  {J.}~\bibnamefont {Billy~et al.}},\ }\href@noop {} {\bibfield  {journal}
  {\bibinfo  {journal} {Nature}\ }\textbf {\bibinfo {volume} {453}},\ \bibinfo
  {pages} {891} (\bibinfo {year} {2008})}\BibitemShut {NoStop}%
\bibitem [{\citenamefont {Roati~et al.}(2008)}]{roati08}%
  \BibitemOpen
  \bibfield  {author} {\bibinfo {author} {\bibfnamefont {G.}~\bibnamefont
  {Roati~et al.}},\ }\href@noop {} {\bibfield  {journal} {\bibinfo  {journal}
  {Nature}\ }\textbf {\bibinfo {volume} {453}},\ \bibinfo {pages} {895}
  (\bibinfo {year} {2008})}\BibitemShut {NoStop}%
\bibitem [{\citenamefont {Das~Sarma}\ \emph {et~al.}(1986)\citenamefont
  {Das~Sarma}, \citenamefont {Kobayashi},\ and\ \citenamefont
  {Prange}}]{sankarprl86}%
  \BibitemOpen
  \bibfield  {author} {\bibinfo {author} {\bibfnamefont {S.}~\bibnamefont
  {Das~Sarma}}, \bibinfo {author} {\bibfnamefont {A.}~\bibnamefont
  {Kobayashi}}, \ and\ \bibinfo {author} {\bibfnamefont {R.~E.}\ \bibnamefont
  {Prange}},\ }\href {\doibase 10.1103/PhysRevLett.56.1280} {\bibfield
  {journal} {\bibinfo  {journal} {Phys. Rev. Lett.}\ }\textbf {\bibinfo
  {volume} {56}},\ \bibinfo {pages} {1280} (\bibinfo {year}
  {1986})}\BibitemShut {NoStop}%
\bibitem [{\citenamefont {Merlin}\ \emph {et~al.}(1985)\citenamefont {Merlin},
  \citenamefont {Bajema}, \citenamefont {Clarke}, \citenamefont {Juang},\ and\
  \citenamefont {Bhattacharya}}]{merlin85}%
  \BibitemOpen
  \bibfield  {author} {\bibinfo {author} {\bibfnamefont {R.}~\bibnamefont
  {Merlin}}, \bibinfo {author} {\bibfnamefont {K.}~\bibnamefont {Bajema}},
  \bibinfo {author} {\bibfnamefont {R.}~\bibnamefont {Clarke}}, \bibinfo
  {author} {\bibfnamefont {F.~Y.}\ \bibnamefont {Juang}}, \ and\ \bibinfo
  {author} {\bibfnamefont {P.~K.}\ \bibnamefont {Bhattacharya}},\ }\href
  {\doibase 10.1103/PhysRevLett.55.1768} {\bibfield  {journal} {\bibinfo
  {journal} {Phys. Rev. Lett.}\ }\textbf {\bibinfo {volume} {55}},\ \bibinfo
  {pages} {1768} (\bibinfo {year} {1985})}\BibitemShut {NoStop}%
\bibitem [{\citenamefont {Vega}\ \emph {et~al.}(2010)\citenamefont {Vega},
  \citenamefont {Conde}, \citenamefont {McBride}, \citenamefont {Abascal},
  \citenamefont {Noya}, \citenamefont {Ramirez},\ and\ \citenamefont
  {Ses\'{e}}}]{vega10}%
  \BibitemOpen
  \bibfield  {author} {\bibinfo {author} {\bibfnamefont {C.}~\bibnamefont
  {Vega}}, \bibinfo {author} {\bibfnamefont {M.~M.}\ \bibnamefont {Conde}},
  \bibinfo {author} {\bibfnamefont {C.}~\bibnamefont {McBride}}, \bibinfo
  {author} {\bibfnamefont {J.~L.~F.}\ \bibnamefont {Abascal}}, \bibinfo
  {author} {\bibfnamefont {E.~G.}\ \bibnamefont {Noya}}, \bibinfo {author}
  {\bibfnamefont {R.}~\bibnamefont {Ramirez}}, \ and\ \bibinfo {author}
  {\bibfnamefont {L.~M.}\ \bibnamefont {Ses\'{e}}},\ }\href {\doibase
  10.1063/1.3298879} {\bibfield  {journal} {\bibinfo  {journal} {The Journal of
  Chemical Physics}\ }\textbf {\bibinfo {volume} {132}},\ \bibinfo {eid}
  {046101} (\bibinfo {year} {2010})}\BibitemShut {NoStop}%
\bibitem [{\citenamefont {Paesani}\ \emph {et~al.}(2010)\citenamefont
  {Paesani}, \citenamefont {Yoo}, \citenamefont {Bakker},\ and\ \citenamefont
  {Xantheas}}]{paesani10}%
  \BibitemOpen
  \bibfield  {author} {\bibinfo {author} {\bibfnamefont {F.}~\bibnamefont
  {Paesani}}, \bibinfo {author} {\bibfnamefont {S.}~\bibnamefont {Yoo}},
  \bibinfo {author} {\bibfnamefont {H.~J.}\ \bibnamefont {Bakker}}, \ and\
  \bibinfo {author} {\bibfnamefont {S.~S.}\ \bibnamefont {Xantheas}},\ }\href
  {\doibase 10.1021/jz100734w} {\bibfield  {journal} {\bibinfo  {journal} {The
  Journal of Physical Chemistry Letters}\ }\textbf {\bibinfo {volume} {1}},\
  \bibinfo {pages} {2316} (\bibinfo {year} {2010})},\ \Eprint
  {http://arxiv.org/abs/http://pubs.acs.org/doi/pdf/10.1021/jz100734w}
  {http://pubs.acs.org/doi/pdf/10.1021/jz100734w} \BibitemShut {NoStop}%
\bibitem [{\citenamefont {González}\ \emph {et~al.}(2010)\citenamefont
  {González}, \citenamefont {Noya}, \citenamefont {Vega},\ and\ \citenamefont
  {Sesé}}]{briesta10}%
  \BibitemOpen
  \bibfield  {author} {\bibinfo {author} {\bibfnamefont {B.~S.}\ \bibnamefont
  {González}}, \bibinfo {author} {\bibfnamefont {E.~G.}\ \bibnamefont
  {Noya}}, \bibinfo {author} {\bibfnamefont {C.}~\bibnamefont {Vega}}, \ and\
  \bibinfo {author} {\bibfnamefont {L.~M.}\ \bibnamefont {Sesé}},\ }\href
  {\doibase 10.1021/jp910770y} {\bibfield  {journal} {\bibinfo  {journal} {The
  Journal of Physical Chemistry B}\ }\textbf {\bibinfo {volume} {114}},\
  \bibinfo {pages} {2484} (\bibinfo {year} {2010})},\ \bibinfo {note} {pMID:
  20121175},\ \Eprint
  {http://arxiv.org/abs/http://pubs.acs.org/doi/pdf/10.1021/jp910770y}
  {http://pubs.acs.org/doi/pdf/10.1021/jp910770y} \BibitemShut {NoStop}%
\bibitem [{\citenamefont {Li}\ \emph {et~al.}(2011)\citenamefont {Li},
  \citenamefont {Walker},\ and\ \citenamefont {Michaelides}}]{zheng11}%
  \BibitemOpen
  \bibfield  {author} {\bibinfo {author} {\bibfnamefont {X.-Z.}\ \bibnamefont
  {Li}}, \bibinfo {author} {\bibfnamefont {B.}~\bibnamefont {Walker}}, \ and\
  \bibinfo {author} {\bibfnamefont {A.}~\bibnamefont {Michaelides}},\ }\href
  {\doibase 10.1073/pnas.1016653108} {\bibfield  {journal} {\bibinfo  {journal}
  {Proceedings of the National Academy of Sciences}\ }\textbf {\bibinfo
  {volume} {108}},\ \bibinfo {pages} {6369} (\bibinfo {year} {2011})},\ \Eprint
  {http://arxiv.org/abs/http://www.pnas.org/content/108/16/6369.full.pdf+html}
  {http://www.pnas.org/content/108/16/6369.full.pdf+html} \BibitemShut
  {NoStop}%
\end{thebibliography}%

\end{document}